# The EMory BrEast imaging Dataset (EMBED): A Racially Diverse, Granular Dataset of 3.5M Screening and Diagnostic Mammograms

Jiwoong J. Jeong[1], Brianna L. Vey[2], Ananth Reddy[2], Thomas Kim[3], Thiago Santos[1], Ramon Correa[1], Raman Dutt[2], Marina Mosunjac[4], Gabriela Oprea-Ilies[4], Geoffrey Smith[4], Minjae Woo[5], Christopher R. McAdams[2], Mary S. Newell[2], Imon Banerjee[6], Judy Gichoya[2], Hari Trivedi[2]

**Affiliations and Addresses**
[1] Department of Computer Science, Emory University, 201 Dowman Drive, Atlanta, GA, 30322, USA.
[2] Department of Radiology, Emory University, 1364 Clifton Rd, Atlanta, GA, 30322, USA
[3] College of Computing, Georgia Institute of Technology, 801 Atlantic Dr NW, Atlanta, GA 30332, USA
[4] Department of Pathology, Emory University, 1364 Clifton Rd, Atlanta, GA, 30322, USA
[5] School of Data Science and Analytics, Kennesaw State University, 1000 Chastain Road, Kennesaw, GA 30144, USA
[6] Department of Biomedical Informatics, Emory University, 100 Woodruff Circle, Atlanta, GA, 30322, USA.

**Abstract**
Developing and validating artificial intelligence models in medical imaging requires datasets that are large, granular, and diverse. To date, the majority of publicly available breast imaging datasets lack in one or more of these areas. Models trained on these data may therefore underperform on patient populations or pathologies that have not previously been encountered.  The EMory BrEast imaging Dataset (EMBED) addresses these gaps by providing 3650,000 2D and DBT screening and diagnostic mammograms for 116,000 women divided equally between White and African American patients. The dataset also contains 40,000 annotated lesions linked to structured imaging descriptors and 61 ground truth pathologic outcomes grouped into six severity classes. Our goal is to share this dataset with research partners to aid in development and validation of breast AI models that will serve all patients fairly and help decrease bias in medical AI.

**Background**

Breast cancer detection remains one of the most frequent commercial and research applications for deep learning (DL) in radiology[1–3]. Development of DL models to improve breast cancer screening and detection requires robustly curated and demographically diverse datasets to ensure models are generalizable. While many breast cancer datasets have been released publicly, these are largely ethnically and racially homogenous[4–6]. Specially, African American patients are severely underrepresented in breast imaging and other healthcare-related datasets. While a recent large study from Google demonstrated an overall improvement in breast cancer screening recall rates when using their model alongside radiologists, very few African American patients were included in the study[7]. Our institution's population of 45% African American patients puts us in a unique position to extract and curate a diverse dataset to improve representation of African American women in breast imaging research.

Limited datasets lead to brittle AI models[8–10] and results in underperformance on patients not included in training data which can cause inadvertent systemic racial bias and disparities in care[11–14]. For example, many models trained for skin cancer detection and genomics use data consisting of up to 96% Caucasian or European population groups[15,16]. Beyond racial diversity, current breast imaging datasets are lacking in either size or granularity. A large dataset created for the DREAM challenge consists of 640,000 screening mammograms from 86,000 subjects weakly labeled as *benign* or *malignant*, but was composed of <0.2% positive cases [4] and no regions of interest (ROIs). The CBIS-DDSM dataset[6] contains 2,620 scanned film mammograms with lesion annotations, however scanned filmed mammograms are technically different from full field digital mammogram (FFDM) and therefore cannot be used in isolation to train AI models for FFDM. The only large (>10k cases), publicly available dataset that contains lesion level ROIs and stratified pathology diagnoses is the Optimam Mammography Image Database (OMI-DB)[17], however this dataset does not contain semantic imaging descriptors. The EMory BrEast imaging Dataset (EMBED) bridges these gaps of lack of diversity, granularity and scale to allow development of robust AI models for screening mammography with potential for decreased bias in performance for African American patients who usually experience worse outcomes from breast cancer[18,19]. We consolidate imaging, pathology, demographics, and outcomes into a curated anonymized dataset that can be used by AI community to develop DL model for breast cancer screening and detection, validation of existing models, and also support population-level research of outcomes include healthcare inequities.

**Dataset Description**

With the approval of Emory Institutional Review Board (IRB), this retrospective dataset of curated screening and diagnostic mammogram studies was developed using largely automated and semi-automated curation techniques which are detailed below. This was possible because image acquisition, reporting, lexicon, followup, and management for breast imaging is governed by the Breast Imaging, Data and Reporting System (BI-RADS)[20] and the Mammography Quality Standards Act (MQSA) [21]. These standards result in a high level of data homogeneity within and across institutions, easing the process of dataset development and increasing the likelihood that developed models will generalize to outside populations.

*Inclusion Criteria*

We identified patients who had received a screening or diagnostic mammogram at our institution from 2013-2020. Data was collected from four institutional hospitals, including two community hospitals, one large inner-city hospital, and a private academic hospital. Inclusion criteria were age 18 years or older, female gender, and availability of at least one mammogram in our PACS. An overview of the full dataset is provided in Figure 1.

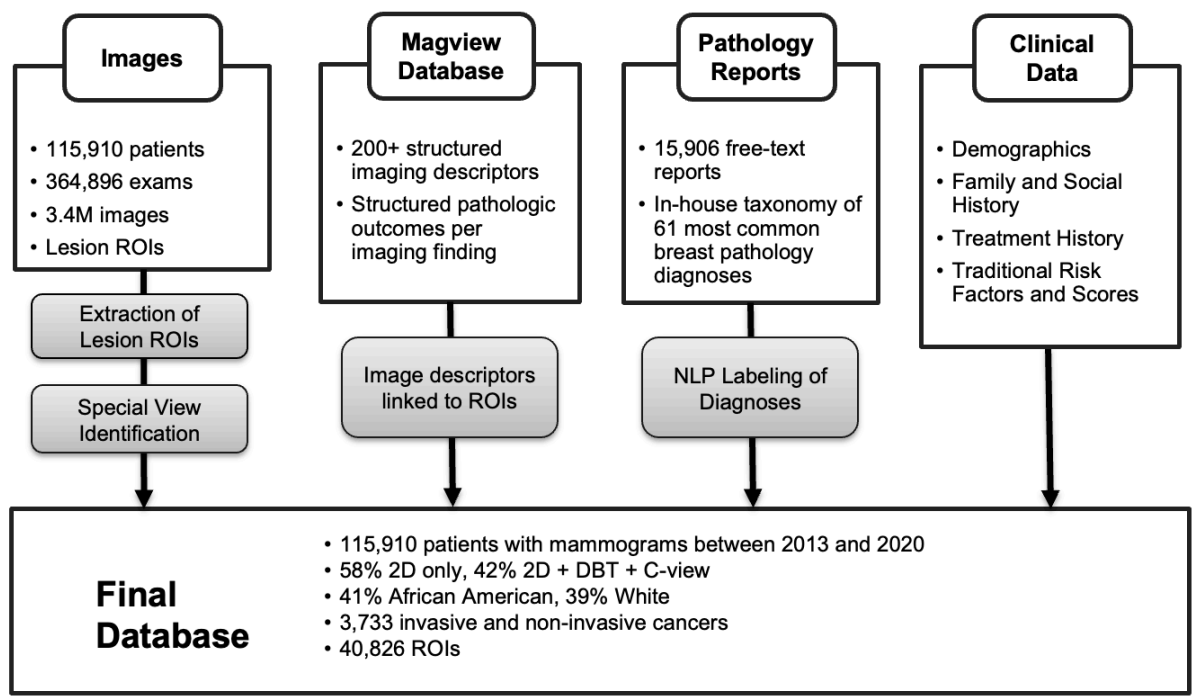

**Figure 1.** Overview of extraction and curation sof the EMBED dataset with four core components: images and ROIs, structured imaging descriptors and pathologic outcomes from MagView, free-text pathology reports, and additional clinical data.

## *Cohort Description*

Patient information is summarized in Table 1. A total of 115,910 unique patients with 364,896 screening and diagnostic mammograms are available. Mean age of patients overall and at first mammogram is 58.5 ± 12.1 years and 55.3 ± 12.2 years and, respectively. Racial distribution is 48,246 (41.6 %) African American, 45,114 (38.9 %) White, 7,552 (6.5%) Asian, 1,130 (1.0%) Native Hawaiian / Pacific Islander, and 13,050 (11.3%) Unknown. Ethnic distribution is 88,025 (75.9%) non-Hispanic, 6,486 (5.6%) Hispanic, and 21,399 (18.5%) Unknown. 37,939 patients have at least 3 years follow-up and 24,933 patients have 5 years follow-up; overall distribution is shown in Figure 2. There are 3,733 (3.2%) total patients with cancers with an annual cancer incidence of 1.16 ± 0.15% on screening mammography.

| Overall Dataset Statistics | |
|---|---|
| Number of patients | 116902 |
| Mean age | 58.5 ± 12.1 |
| Mean age at first visit | 55.3 ± 12.2 |
| Screening Mammograms | 281,509 |
| Diagnostic Mammograms | 83,387 |
| Mean annual recall rate | 10.6 ± 1.6 % |
| **Race** | |
| African American | 48,246 (41.6 %) |
| White | 45,114 (38.9 %) |
| Asian | 7,552 (6.5 %) |
| Native Hawaiian / Pacific Islander | 1,130 (1.0 %) |
| Multiple | 510 (0.4 %) |
| American Indian or Alaskan Native | 308 (0.3 %) |
| Unknown | 13,050 (11.3 %) |
| **Ethnicity** | |
| Hispanic | 6,486 (5.6 %) |
| Non-Hispanic | 88,025 (75.9 %) |
| Unknown | 21,399 (18.5%) |
| **Cancer Prevalence** | |
| Total Cancers | 3733 (3.2%) |
| Annual cancer incidence | 1.16 ± 0.15% |
| **Regions of Interest** | |
| Total regions of interest | 40,826 |
| Regions of interest directly linked to findings | 32,448 |

**Table 1.** Descriptive statistics for the full EMBED dataset which contains approximately even numbers of African American and White patients. Regions of interest are annotated by interpreting radiologists and can be linked directly to a single finding in approximately 80% of cases. The remaining 20% of ROIs arise from cases with multiple findings and require manual linkage.

## Data Extraction

### Images

All mammographic exams were extracted from the institutional PACS in DICOM format using Niffler[22] - an open-source pipeline developed in-house for retrospective image extraction that leverages pydicom[23]. Of 281,509 screening and 83,387 diagnostic exams, 148,320 (52.7%) screening and 65,265 (78.3%) diagnostic exams are 2D only, and 133,189 (47.3%) screening and 18,122 (21.7%) diagnostic exams are 2D + digital breast tomosynthesis (DBT) + synthetic 2D C-view. All images were converted to 16-bit

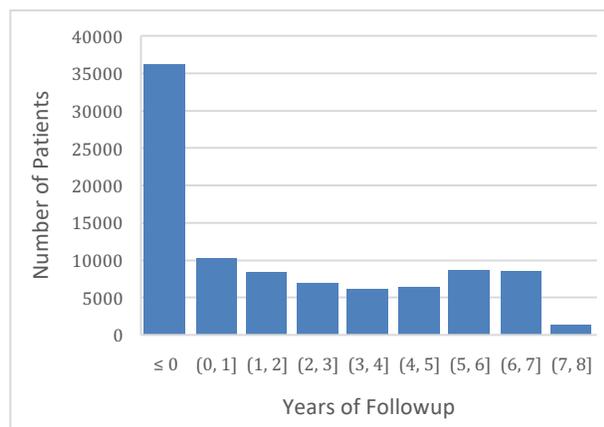

**Figure 2**. Distribution of the follow-up period available per patient. A total of 37,939 patients have at least three years of follow-up and 24,933 have at least 5 years of follow-up.

PNG format and DICOM metadata were extracted and de-identified. Dates were shifted following HIPAA guidelines with fixed patient-level offset key to maintain temporality between multiple imaging exams and associated clinical data. A master key is retained for adding new exams to the dataset at regular intervals. Data anonymization was performed using an open-source in-house python library[24].

*Imaging findings*

Imaging findings are recorded at the time of interpretation in Magview (Fulton, MD) and output into a structured database that includes information such as exam type, reason for visit, BI-RADS score, and BI-RADS imaging descriptors[25]. Imaging descriptors include information on appearance of masses, calcification, distribution and location of findings, presence of implants, and additional non-localized imaging findings such as global asymmetries (Table 2). Findings are notated on a *per finding and per breast* basis, therefore each exam may have zero to several findings for each breast. The distribution of findings type separated by BI-RADS score and exam type are shown in Table 3.

| Encoded Descriptors | Possible Values |
| --- | --- |
| **Mass** | |
| Shape | Generic (G), Round (R), Oval (O), Irregular (X), Questioned architectural distortion (Q), Architectural Distortion (A), Asymmetric tubular structure/solitary dilated duct (T), Intramammary lymph nodes (N), Global asymmetry (B), Focal asymmetry (F), Developing asymmetry (V), Lymph Node (Y) |
| Margin | Circumscribed (D), Obscured (U), Microlobulated (M), Indistinct (I), Spiculated (S) |
| Density | High density (+), Isodense (=), Low density (-), Fat containing (0) |
| **Calcification** | |
| Finding | Amorphous (A), Benign (9), Coarse heterogenous (H), Course Popcorn-like (C), Dystrophic (D), Rim (E), Fine-linear (F), Fine linear-branching (B), Generic (G), Fine Pleomorphic (I), Large rodlike (L), Mild of calcium (M), Oil Cyst (J), Pleomorphic (K), Punctate (P), Round (R), Skin (S), Lucent centered (O), Suture (U), Vascular (V), Coarse (Q) |
| Distribution | Grouped (G), Segmental (S), Regional (R), Diffuse/scattered (D), Linear (L), Clustered (C) |
| **Other** | |
| Size and Position | Side (L or R), Size (in mm), Location (quadrant, sub-areolar, or axillary tail), Depth (anterior, middle, posterior), Distance (in cm) |
| Related Findings to Primary Finding | Post-lumpectomy change (U), Post-lumpectomy and radiation change (1), Post-surgical change (P), Biopsy Clip (W), Post-reduction change (C), Focal Asymmetry (Q), Asymmetry (4), Prominent lymph node (2), Mastectomy and flap reconstruction (!) |

**Table 2**. Samples of the imaging descriptor categories and commonly encoded values available from MagView. Each finding described by the radiologist at the time of interpretation is encoded into a structured format. Because the majority of screening exams are negative, most exams have no associated descriptors.

| BIRADS | Screening | Diagnostic | Mass | Calcification | Asymmetry | Architectural Distortion |
|---|---|---|---|---|---|---|
| 0 | 34,943 | - | 6,523 | 7728 | 23,147 | 2,312 |
| 1 | 167,174 | 13,776 | 5 | 0 | 132 | 16 |
| 2 | 23,289 | 25,960 | 10,562 | 7,666 | 8,243 | 434 |
| 3 | - | 17,053 | 2,915 | 6,106 | 5,431 | 275 |
| 4 | - | 6,860 | 2742 | 4,261 | 2,184 | 690 |
| 5 | - | 649 | 911 | 342 | 123 | 60 |
| 6 | - | 1,088 | 659 | 249 | 208 | 52 |

**Table 3.** A sample of imaging findings for the training and validation dataset, categorized broadly by masses, calcifications, asymmetries, and architectural distortion. More detailed findings information is available as shown in Table 2. Information regarding findings distributions in the test set is withheld.

*Pathologic Outcomes*
Pathologic outcomes are clinically correlated and manually recorded in Magview on a *per finding* basis by administrative staff. This includes results of fine need aspiration (FNA), biopsy, lumpectomy, and mastectomy for breast tissue or lymph nodes. Each finding may contain up to 10 associated pathology results, but rarely contains more than 4. We also separately extracted all free-text pathology records from an institutional database (CoPath, Pittsburgh, PA) for troubleshooting in case of discrepancies in Magview results. Pathology results outside the breast, chest wall, or axilla are excluded, even if the primary malignancy is breast (e.g., a lung metastasis from primary breast cancer).

*Demographics, Family, and Treatment History*
Patient demographics including age, race, ethnicity, and insurance status are collected for each patient. Family, procedure, and treatment history, including hormone replacement therapy are available for many patients but subject to the information being recorded at the time of patient visit. Traditional risk factors and Gail and Tyrer-Cuzick risk scores are collected, when available.

**Training, Validation, and Test Sets**
We have randomly divided the data into 80% training and validation and 20% test sets at the patient level. All exams for a given patient are included in the respective set and there is no patient leakage between sets. Descriptive statistics are provided for the training and validation sets but withheld for the test set which is reserved internally for model validation. Institutions may divide the training and validation data as needed; however, it is currently divided into 8 equal patient cohorts containing approximately 11,000 patients each.

**Data Curation**

*Imaging*

There were four challenges for image data curation: 1) differentiation of 2D, digital breast tomosynthesis (DBT), and C-view images, 2) differentiation between standard mammographic views and special views (spot, magnification, or procedural views) in diagnostic exams, 3) extraction of burned-in regions of interest (ROIs) that are saved directly into pixel data on a copy of the original mammogram (Screensave image), and 4) extraction of breast tissue from inside of spot compression or magnification paddles on diagnostic exams. We designed a semi-automated supervised machine learning pipeline to overcome these challenges that combines traditional computer vision and deep learning techniques, summarized in Figure 3.

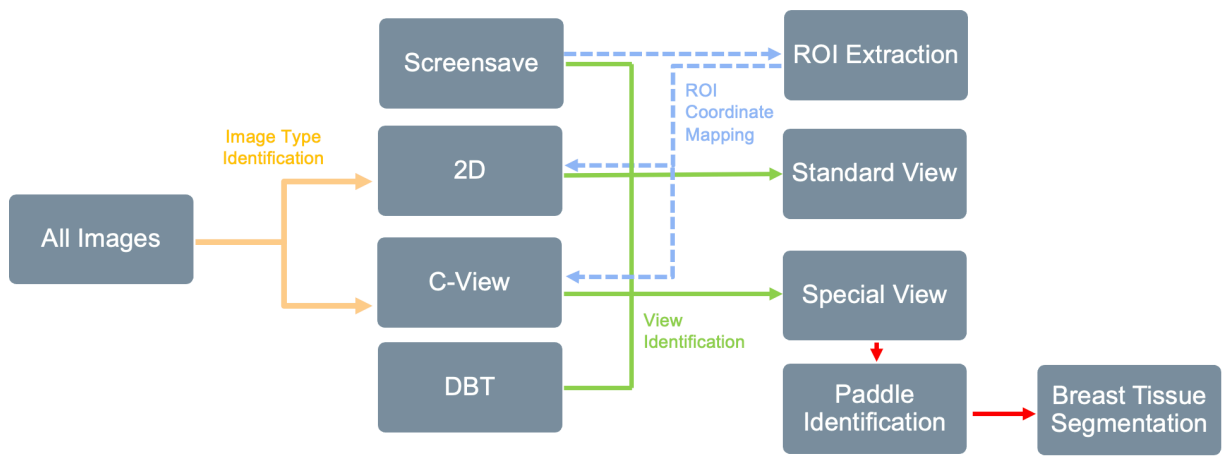

**Figure 3.** Overview of image filtration for classifying image types and extracting relevant regions of interest and tissue patches. This is achieved using a combination of computer vision (CV) techniques, DICOM metadata, and rules-based heuristics. ROI extraction is achieved by identifying ROIs within Screensave images, extracting the ROI coordinates, identifying the matching source mammogram, and then mapping the ROI coordinates back to the original image. Examples of ROI extraction and special view tissue segmentation are provided in figures 4 and 5.

1. Differentiation of Image Type

Approximately 42% of our exams are combined 2D and DBT exams, and therefore can contain up to four image types: 2D, DBT, synthetic 2D (C-view), or Screensave images containing ROIs. We utilized a rules-based approach using the 'SeriesDescription', 'ViewPosition', and 'ImageLaterality' tags in the DICOM metadata to identify and label each image type. Results were manually verified on a random test set of 5,000 images and were 100% accurate.

2. Differentiation between Standard and Special Views

To differentiate between standard and special views for 2D images, supervised image classification and metadata filtration methods were both attempted. To classify based on image appearance alone, a VGG11 deep learning model[26] with batch normalization was trained, tested, and validated on a manually curated 5,000 image dataset containing FFDM and special views, achieving an overall accuracy of 98.46%. Because the results were imperfect, we also examined the DICOM metadata of FFDM and special views and found that a private tag

'0_ViewCodeSequence_0_ViewModifierCodeSequence_CodeMeaning' could be used to differentiate between FFDM and special views. We manually verified the results on a separate test set of 5,000 randomly selected images and confirmed the results were 100% accurate. It is possible that this metadata tag is valid only at our institution, so the image-based classification model is available for public use[27] as it may be more generalizable. Using this technique, we identified 208,254 images containing special views.

3. ROI Extraction and Mapping to Original Mammogram

To automate a method of ROI detection and localization, 450 Screensave images containing ROIs were randomly selected and the location of circular ROIs was annotated by a trained student (JJ) using bounding boxes on the Md.ai platform. Annotations were used were used to train a deep learning object detection model using the EfficientDet-b0 architecture[28] to localize ROIs. Detection accuracy on the test set was 99.99% with an intersection over union (IOU) of .95, including cases with multiple ROIs on an image. We then ran inference for detecting ROIs on all remaining Screensave images and manually verified localization accuracy on a test set of 5,000 cases.

Screensave images are lower-resolution duplicates of an original mammogram and are often flipped horizontally due to the hanging protocol of the interpreting radiologist. To map the extracted ROI from the Screensave image back to the original mammogram, we first identified the original mammogram using an image similarity function from the Simple ITK python library[29]. Once a source mammogram was identified, the ROI coordinates from the Screensave were scaled and translated back to the original mammogram and saved into the database (Figure 4). This process was manually verified on 2,000 image matches to ensure proper matching of the Screensave to the source image and proper translation of the ROI coordinates, with only 3 matching errors discovered(< 0.2%) in which Screensaves were mapped to a C-view instead of 2D image. Approximately 90% of ROIs are from screening exams assigned a BI-RADS 0 assessment, and approximately 70% are sourced from 2D images and 30% from C-view images.

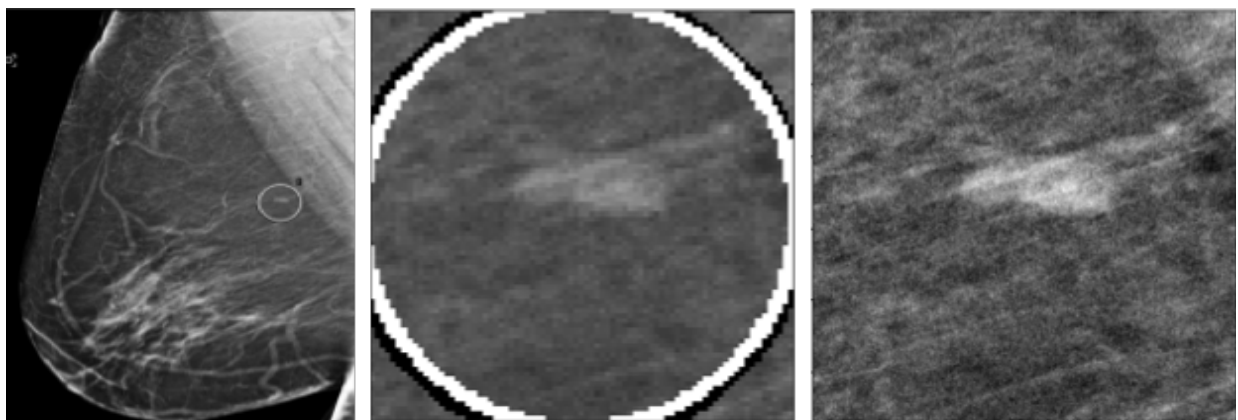

**Figure 4:** Sample of ROI extraction. *Left*: Low resolution Screensave copy of mammogram containing a burned-in ROI annotated by the interpreting radiologist. *Middle*: DL based ROI detection and coordinate extraction from the Screensave image. *Right*: The matching original mammogram is found using image comparison from the the Python Simple ITK library, and the coordinates of the ROI are mapped to the original image.

Because 2D and DBT acquisitions are obtained sequentially during the same visit, the anatomy and tissue distribution is similar between both acquisitions yielding a similar appearance between the 2D and C-view images. Therefore, we increased the number of ROIs by *secondarily* matching each Screensave to the corresponding C-view or 2D image of the source image. For example, if a Screensave's primary match was a 2D left mediolateral oblique (LMO) image, we map its ROI coordinates as *primary* to the 2D left MLO image and as *secondary* to the C-View left MLO image from the same exam. This increased the number of ROIs available by approximately 40%. Because this process is susceptible to errors from variations in patient positioning between acquisitions, secondary ROIs are recorded separately to allow users to include or exclude them during model development. Using these techniques, we extracted and mapped 40,826 ROIs.

Finally, mapping of ROIs to imaging descriptors and pathologic outcomes from Magview can only be done automatically if there is only one finding for the same breast inside of the Magview database. If more than one finding is recorded per breast in an exam, the linkage between the ROI and structured Magview data becomes ambiguous. We differentiate these 'automatically linkable' ROIs from 'ambiguous' ROIs in our dataset. Automatically linkable ROIs have a definite set of imaging descriptors and/or pathologic outcomes from Magview that can be used as a label for the ROI during model training or validation. Ambiguous ROIs would require manual review to link them to the appropriate imaging descriptors inside of Magview. Across all 10 cohorts, 32,448 ROIs are automatically linkable whereas 8,378 are ambiguous.

4. Extraction of Tissue Inside Paddles for Special Diagnostic Views

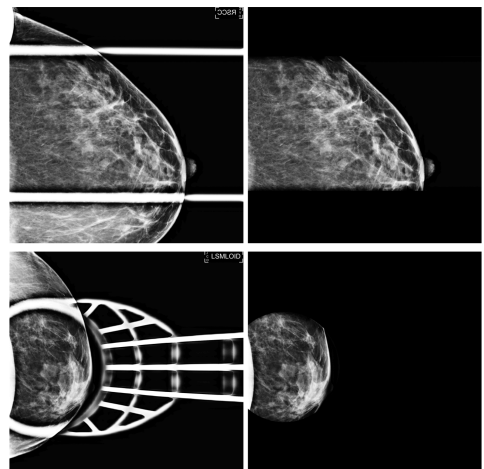

12 distinct paddle types were identified from images with special views. The paddles were categorized according to their shapes, which may be rectangular or circular. Because all paddles are metallic, their pixel intensities were significantly higher than the surrounding tissue. For rectangular paddles, simple thresholding was used to convert the grayscale images into binary images, and a deterministic algorithm was used to identify paddled edges based on a row and column sum of pixels (Figure 5). For the circular paddles, a feature engineering technique, Hough Circle transformation, was utilized to detect any metallic device in a circular shape. A subsequent process was implemented to maximize the size of the radius while leaving the metallic device out of the cropped area.

**Figure 5.** Examples of special magnification and spot compression views (left) and resultant extracted images of tissue inside the paddle (right) achieved using histogram analysis. Extracted tissue is saved as a pixel mask corresponding to the original mammogram.

*Pathology*
Pathology data was curated using in-house taxonomy for breast pathologies (Figure 6) which was created in consultation with breast pathologists and oncologists to identify the 61 most-common findings in breast pathology reports. These were divided into the following seven severity groups:

invasive breast cancer (IBC), in-situ cancer (ISC), high-risk lesion (HRL), borderline lesion (BLL), non-breast cancer (NBC), benign (B), or negative (N).

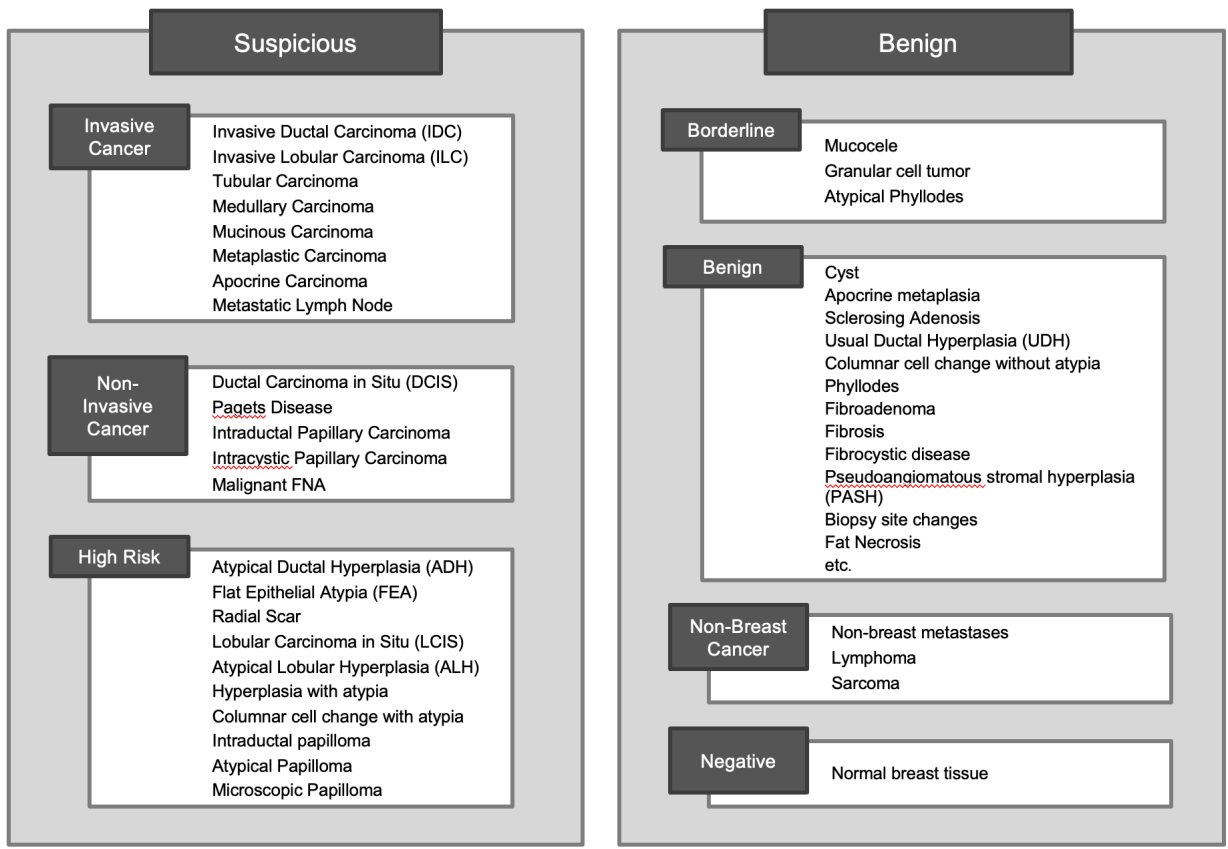

**Figure 6.** Taxonomy of the 61 most-common breast pathologic findings at our institution grouped into 7 categories by severity. Each pathology report is tagged with one or more of diagnoses from this list. The Benign category contains 23 diagnoses but is abbreviated in this table.

Inside Magview, pathology information for each finding was converted to one of the 7 severity groups using a lookup table. For example, if the pathologic outcome for a finding was [ductal carcinoma in situ, flat epithelial atypia], these were encoded as [ISC, HRL]. A second field recorded the *worst* pathologic diagnosis for that finding, which is ISC. The number of ROIs linked to each pathology severity group in screening and diagnostic exams is shown in Table 4.

Because pathology information is encoded into Magview manually by administrative staff, it is occasionally subject to human error. To create a secondary check for pathology outcomes, we trained a hierarchical hybrid natural language processing (NLP) system using 8,000 expert-annotated labels to extract pathologic diagnoses from free-text pathology reports. In the first level of the hierarchy, a transformer-based Character Level BERT[30] was trained to classify one or many of the seven pathologic groups, with an overall F1 score of 94.6% and 96.8% for classifying all pathologic groups and only the worst pathologic group, respectively. Subsequently, at the second level of the hierarchy, six independent discriminators are trained to classify individual pathologic diagnoses with an overall F1 score of 90.7%. The Negative group contains no individual diagnoses so no secondary discriminator is required for this class. The output from this model can be compared directly against the structured pathologic outcomes in Magview and mismatches can be flagged for human review, however the efficacy of this strategy remains to be seen. The model is currently being validated on external datasets and will be published and released separately to allow other institutions to automatically extract diagnoses from breast pathology reports.

| Number of Patients by Category | | Total Screening ROIs | Total Diagnostic ROIs | ROIs Linked Directly to Finding |
|---|---|---|---|---|
| All ROIs | 32,514 | 29,968 | 2,538 | 25,873 |
| Invasive Breast Cancer | 1,765 | 1,383 | 382 | 1,130 |
| In Situ Cancer | 845 | 687 | 158 | 602 |
| High Risk Lesion | 1,146 | 849 | 297 | 778 |
| Borderline Lesion | 24 | 24 | 0 | 16 |
| Benign | 3,281 | 2,625 | 656 | 2,289 |
| Non-breast Cancer | 46 | 19 | 27 | 12 |

**Table 4**. ROI counts by pathologic outcome for the training and validation datasets. Approximately 80% of ROIs can be directly linked to imaging findings and pathologic outcomes. The two commonest ROIs are for benign findings followed by invasive cancers. Information regarding pathologic findings and ROIs in the test set is withheld.

**Database Size and Structure**

The total dataset size is 16.0 terabytes. The Magview data, image metadata, ROI information, and clinical data is stored in a MongoDB[31] database. Each data cohort is a collection in the database, and each document in the collection will represent a datapoint. The documents store data in a key-value pair (JSON) format. Anonymized accession number and Cohort ID are combined to form the primary key. The dynamic schema of MongoDb allows us to store different data attributes in the same collection. Images are stored as 16-bit PNG in a hierarchical folder structure by cohort, patient, exam, and then image. Filenames are hashed so each filename is unique and can be linked directly to its DICOM metadata.

**Challenges**
Throughout the process of collecting, organizing, anonymizing and consolidating the datasets, there were several challenges which required creative solutions and may suggest areas for future innovation to allow for easier adoption at other institutions.

DICOM metadata extraction was designed such that each row in the resultant dataframe represents a single image and each column represents a metadata element and its value. However, DICOM metadata varies across manufacturer and model, and was sometimes corrupted resulting nested values that generated over 2,000 metadata tags for a single file. This caused an explosion of the dataframe, so the decision was made to retain metadata that was present in at least 10% of files. Private metadata fields were dropped. This threshold is a customizable parameter during metadata extraction inside Niffler.

We encountered an obstacle during DICOM to PNG conversion in which some images appeared low-contrast or washed out. This was due to a difference in window-level mapping between General Electric and Hologic scanners. To address this, the Manufacturer DICOM tag was read during PNG conversion. Hologic was converted using min-max windowing and GE was converted by applying the VOI LUT function provided in each image's metadata. If replicating this process at another institution, care should be taken that images from different manufacturers are normalized appropriately by either using built-in normalization functions of pydicom[23], dcmtk[32], or numpy[33].

Lastly, an ongoing challenge remains linkage of ROIs back to imaging and pathologic findings in the Magview database. In exams where there is only a single described finding per breast, one or multiple ROIs can be automatically mapped to this finding. However, when there are multiple findings per breast, ROIs cannot be automatically mapped. These will require development of a new heuristic, which may include automatic selection of ROI based on coded information in Magview for the breast quadrant and depth, or by manual review.

**Data Availability**
To date, EMBED has been used in two validation studies for breast cancer risk prediction[34,35]. Permission for external collaboration are reviewed on a case-by-case basis by the IRB. An IRB is also pending for release of 1000 patients (approximately 3,500 exams), up-sampled for 200 cancers, which will provide researchers with an opportunity to review the structure and content of EMBED before deciding whether to carry out an analysis on the full dataset.

**Support**
This work was partially supported the National Center for Advancing Translational Sciences of the National Institutes of Health under Award Number UL1TR002378.

Bibliography

1. Tadavarthi Y, Vey B, Krupinski E, et al. The state of radiology AI: considerations for purchase decisions and current market offerings. *Radiol Artif Intell*. 2020;2(6):e200004. doi:10.1148/ryai.2020200004
2. Houssami N, Kirkpatrick-Jones G, Noguchi N, Lee CI. Artificial Intelligence (AI) for the early detection of breast cancer: a scoping review to assess AI's potential in breast screening practice. *Expert Rev Med Devices*. 2019;16(5):351-362. doi:10.1080/17434440.2019.1610387
3. Batchu S, Liu F, Amireh A, Waller J, Umair M. A review of applications of machine learning in mammography and future challenges. *Oncology*. 2021;99(8):483-490. doi:10.1159/000515698
4. Schaffter T, Buist DSM, Lee CI, et al. Evaluation of combined artificial intelligence and radiologist assessment to interpret screening mammograms. *JAMA Netw Open*. 2020;3(3):e200265. doi:10.1001/jamanetworkopen.2020.0265
5. Halling-Brown MD, Warren LM, Ward D, et al. OPTIMAM Mammography Image Database: a large scale resource of mammography images and clinical data. *arXiv*. April 2020.
6. Lee RS, Gimenez F, Hoogi A, Miyake KK, Gorovoy M, Rubin DL. A curated mammography data set for use in computer-aided detection and diagnosis research. *Sci Data*. 2017;4:170177. doi:10.1038/sdata.2017.177
7. McKinney SM, Sieniek M, Godbole V, et al. International evaluation of an AI system for breast cancer screening. *Nature*. 2020;577(7788):89-94. doi:10.1038/s41586-019-1799-6
8. Zech JR, Badgeley MA, Liu M, Costa AB, Titano JJ, Oermann EK. Variable generalization performance of a deep learning model to detect pneumonia in chest radiographs: A cross-sectional study. *PLoS Med*. 2018;15(11):e1002683. doi:10.1371/journal.pmed.1002683
9. Pan I, Agarwal S, Merck D. Generalizable Inter-Institutional Classification of Abnormal Chest Radiographs Using Efficient Convolutional Neural Networks. *J Digit Imaging*. 2019;32(5):888-896. doi:10.1007/s10278-019-00180-9
10. Thrall JH, Li X, Li Q, et al. Artificial intelligence and machine learning in radiology: opportunities, challenges, pitfalls, and criteria for success. *J Am Coll Radiol*. 2018;15(3 Pt B):504-508. doi:10.1016/j.jacr.2017.12.026
11. Obermeyer Z, Powers B, Vogeli C, Mullainathan S. Dissecting Racial Bias in an Algorithm used to Manage the Health of Populations. *Science*. 2019;366(6464):447-453. doi:10.1126/science.aax2342
12. Noor P. Can we trust AI not to further embed racial bias and prejudice? *BMJ*. 2020;368:m363. doi:10.1136/bmj.m363
13. Aizer AA, Wilhite TJ, Chen M-H, et al. Lack of reduction in racial disparities in cancer-specific mortality over a 20-year period. *Cancer*. 2014;120(10):1532-1539. doi:10.1002/cncr.28617
14. Gianfrancesco MA, Tamang S, Yazdany J, Schmajuk G. Potential biases in machine learning algorithms using electronic health record data. *JAMA Intern Med*. 2018;178(11):1544-1547. doi:10.1001/jamainternmed.2018.3763
15. Esteva A, Kuprel B, Novoa RA, et al. Dermatologist-level classification of skin cancer with

deep neural networks. *Nature*. 2017;542(7639):115-118. doi:10.1038/nature21056

16. Bustamante CD, Burchard EG, De la Vega FM. Genomics for the world. *Nature*. 2011;475(7355):163-165. doi:10.1038/475163a
17. Halling-Brown MD, Looney PT, Patel MN, Warren LM, Mackenzie A, Young KC. The oncology medical image database (OMI-DB). In: Law MY, Cook TS, eds. *Medical Imaging 2014: PACS and Imaging Informatics: Next Generation and Innovations*. Vol 9039. SPIE Proceedings. SPIE; 2014:903906. doi:10.1117/12.2041674
18. Yedjou CG, Sims JN, Miele L, et al. Health and racial disparity in breast cancer. *Adv Exp Med Biol*. 2019;1152:31-49. doi:10.1007/978-3-030-20301-6_3
19. Newman LA, Kaljee LM. Health Disparities and Triple-Negative Breast Cancer in African American Women: A Review. *JAMA Surg*. 2017;152(5):485-493. doi:10.1001/jamasurg.2017.0005
20. Sickles, EA, D'Orsi CJ, Bassett LW, et al. *ACR BI-RADS Mammography.* American College of Radiology; 2013.
21. Mammography Quality Standards Act and Program | FDA. https://www.fda.gov/radiation-emitting-products/mammography-quality-standards-act-and-program. Accessed August 13, 2021.
22. Kathiravelu P, Sharma P, Sharma A, et al. A DICOM Framework for Machine Learning and Processing Pipelines Against Real-time Radiology Images. *J Digit Imaging*. 2021;34(4):1005-1013. doi:10.1007/s10278-021-00491-w
23. Mason D. SU-E-T-33: Pydicom: An Open Source DICOM Library. *Med Phys*. 2011;38(6Part10):3493-3493. doi:10.1118/1.3611983
24. HITI-anon-internal · TestPyPI. https://test.pypi.org/project/HITI-anon-internal/. Accessed January 7, 2022.
25. Shikhman R, Keppke AL. Breast, Imaging, Reporting and Data System (BI RADS). *StatPearls*. January 2018.
26. Simonyan K, Zisserman A. Very Deep Convolutional Networks for Large-Scale Image Recognition. *arxiv.org*. 2014;cs.CV.
27. GitHub - EMBED Codebase. https://github.com/Emory-HITI/Mammo. Accessed February 6, 2022.
28. Tan M, Pang R, Le QV. Efficientdet: scalable and efficient object detection. In: *2020 IEEE/CVF Conference on Computer Vision and Pattern Recognition (CVPR)*. IEEE; 2020:10778-10787. doi:10.1109/CVPR42600.2020.01079
29. Beare R, Lowekamp B, Yaniv Z. Image Segmentation, Registration and Characterization in R with SimpleITK. *J Stat Softw*. 2018;86. doi:10.18637/jss.v086.i08
30. El Boukkouri H, Ferret O, Lavergne T, Noji H, Zweigenbaum P, Tsujii J. CharacterBERT: Reconciling ELMo and BERT for Word-Level Open-Vocabulary Representations From Characters. In: *Proceedings of the 28th International Conference on Computational Linguistics*. Stroudsburg, PA, USA: International Committee on Computational Linguistics; 2020:6903-6915. doi:10.18653/v1/2020.coling-main.609
31. Plugge E, Membrey P, Hawkins T. Introduction to MongoDB. In: *The Definitive Guide to Mongodb*. Berkeley, CA: Apress; 2010:3-17. doi:10.1007/978-1-4302-3052-6_1
32. dicom.offis.de - DICOM Software made by OFFIS - DCMTK - DICOM Toolkit. https://dicom.offis.de/dcmtk.php.en. Accessed January 10, 2022.

33. Oliphant TE. *Guide to NumPy*. CreateSpace Independent Publishing Platform; 2015.
34. Yala A, Mikhael PG, Strand F, et al. Multi-Institutional Validation of a Mammography-Based Breast Cancer Risk Model. *J Clin Oncol*. November 2021:JCO2101337. doi:10.1200/JCO.21.01337
35. Yala A, Mikhael PG, Lehman C, et al. Optimizing risk-based breast cancer screening policies with reinforcement learning. *Nat Med*. January 2022. doi:10.1038/s41591-021-01599-w